\documentclass[12pt]{article}
\usepackage[left=2.5cm,top=2.50cm,right=2.5cm,bottom=2.50cm]{geometry}
\usepackage{mathrsfs}
\usepackage{amsmath,amssymb,latexsym,color,cancel,graphicx,bbm,colortbl}
\usepackage[english]{babel}
\usepackage[latin1]{inputenc}
\usepackage{ragged2e}
\usepackage{cite}
\usepackage{subfig}
\begin{document}
\date{}
\title{The Generalized Fokker-Planck Equation in terms of Dunkl-type Derivatives}
\author{R. D. Mota$^{a}$,   D. Ojeda-Guill\'en$^{b}$\footnote{{\it E-mail address:} dojedag@ipn.mx} and  M. A. Xicot\'encatl$^{c}$}
\maketitle

\begin{minipage}{0.9\textwidth}
\small $^{a}$ Escuela Superior de Ingenier{\'i}a Mec\'anica y El\'ectrica, Unidad Culhuac\'an,
Instituto Polit\'ecnico Nacional, Av. Santa Ana No. 1000, Col. San
Francisco Culhuac\'an, Alc. Coyoac\'an, C.P. 04430, Ciudad de M\'exico, Mexico.\\

\small $^{b}$ Escuela Superior de C\'omputo, Instituto Polit\'ecnico Nacional,
Av. Juan de Dios B\'atiz esq. Av. Miguel Oth\'on de Mendiz\'abal, Col. Lindavista,
Alc. Gustavo A. Madero, C.P. 07738, Ciudad de M\'exico, Mexico.\\

\small $^{c}$ Departamento de Matem\'aticas, Centro de Investigaci\'on y  Estudios Avanzados del IPN, C.P. 07360, Ciudad de M\'exico, Mexico.\\
\end{minipage}

\begin{abstract}
In this work we introduce two different generalizations of the Fokker-Planck equation in (1+1) dimensions by replacing the spatial derivatives in terms of generalized Dunkl-type derivatives involving reflection operators. As applications of these results, we solve exactly the generalized Fokker-Planck equations for the harmonic oscillator and the centrifugal-type potentials.

\end{abstract}


\section{Introduction}
Fokker and Planck were the first to derive a differential equation describing the probability distribution function of Brownian motion and its fluctuations. Ever since then, the so-called Fokker-Planck equation has been applied to treat a wide variety of problems in physics and its solutions have been obtained using different approaches, including analytical and numerical methods \cite{risken,libro1,libro2,libro3,junker}. Furthermore, the transformation of the Fokker-Planck equation to a Schr\"odinger-type equation was an interesting achievement that allowed all known methods of quantum mechanics to be applied to find its solutions. Among them are  group theory, supersymmetric quantum mechanics and shape invariance \cite{gt1,gt2,junker,prl,polloto,anjos}.

On the other hand, derivative operators involving reflections were first introduced by Wigner \cite{wigner} and later applied by Yang to the harmonic oscillator \cite{YANG}. Independently, Dunkl operators were introduced in \cite{dunkl} as a generalization of the directional derivative. These derivatives can be used to generalize the standard momentum operator, which then continues to satisfy the fundamental quantum-mechanical commutation relations and obeys the classical equations of motion.

The Dunkl derivative, which involves reflection operators, was applied in quantum mechanics to study some physical problems such as the harmonic oscillator and the Coulomb potential in order to obtain their exact solutions, superintegrability and symmetry algebra \cite{GEN1,GEN2,GEN3,GEN4}. In these works the standard derivatives in the Schr\"odinger equation are replaced by the Dunkl derivatives depending on parameters that may be helpful to adjust the theoretical results to better match the experimental results. The applications of the Dunkl derivative were extended to study and solve relativistic and non-relativistic quantum mechanical problems \cite{NOS1,NOS2,NOS3,NOS4,GAZ1,GAZ2,GAZ3,SCH,SCH2}. The Dunkl derivative has also been applied to study statistical and thermodynamic properties of physical problems \cite{DONGT,HAM1,HAM2,BER}. Even more, the Dunkl derivative remains an active field of research, as can be seen in the references \cite{SH1,SH2,SH3,SH4}.

In addition to the Dunkl derivative, other Dunkl-type derivatives involving reflection operators with one, two or three parameters were introduced in relation to some quantum mechanical systems \cite{chung1,chung2}. The Dunkl-type operators also generalize the momentum operator while the fundamental quantum-mechanical commutation relations and the classical equations of motion remain fulfilled. This is the motivation why these Dunkl-type derivatives can be used to study other physical differential equations.

In this direction, the aim of the present work is to generalize the Fokker-Planck equation in terms of two of these generalized Dunkl-type derivatives. As applications of such, we solve exactly the centrifugal-type potential $a(a-1)/x^2$ and the harmonic oscillator plus a centrifugal-type potential.

The paper is organized as follows. In Section 2, we obtain the Schr\"odinger equation of supersymmetric quantum mechanics from the Fokker-Planck equation and identify the drift potential with the superpotential of supersymmetric quantum mechanics. Then, we introduce the Yang, Dunkl and the Chung-Hassanabadi derivatives, which will allow us to generalize the Fokker-Planck equation with two different Dunkl-type derivatives depending on two parameters. The eigenfunctions and energy spectrum of the potential $a(a-1)/x^2$ are obtained in Section 3 for the generalized Fokker-Planck equation in terms of the Chung-Hassanabadi derivative. Similarly, Section 4 is dedicated to solve exactly the generalized Fokker-Planck equation for the harmonic oscillator plus a centrifugal-type potential in terms of a two-parameter Dunkl-type derivative. Finally, we give some concluding remarks in Section 5.

\section{The generalized Fokker-Planck equation by a two-parameter Dunkl-type derivative}

The Fokker-Planck equation in one dimension for the probability density $\mathcal{P}(x,t)$ is
\begin{equation}
\frac{\partial \mathcal{P}(x,t)}{\partial t} =\left(-\frac{\partial }{\partial x}D^{(1)}(x)+\frac{\partial^2}{\partial x^2} D^{(2)}(x)\right)\mathcal{P}(x,t). \label{FP}
\end{equation}
In this expression $D^{(1)}(x)$ and $D^{(2)}(x)$ are the drift and diffusion functions, respectively. The diffusion function is usually set to be a
constant \cite{polloto,anjos,englefield,ho}, which leads us to propose $D^{(2)}(x)=1$ and the drift coefficient as
\begin{equation}
 D^{(1)}(x)=2w(x) \label{driftsuper}.
\end{equation}
If we consider the probability density to be given by
\begin{equation}
\mathcal{P}(x,t)= e^{-\lambda t}e^{\int w(x)dx}\phi(x),
\end{equation}
we arrive to an equation formally identical to Schr\"odinger equation
\begin{equation}
H\phi(x)\equiv \left(-\frac{d^2 }{d x^2}+w(x)^2+w'(x)\right)\phi(x)=\lambda\phi(x).  \label{susy}
\end{equation}
Here, the eigenvalues equation is written in terms of the supersymmetric quantum mechanics superpotential $w(x)$. Thus, the Fokker-Planck equation can be studied using the theory and results of the standard supersymmetric quantum mechanics \cite{risken,prl,junker,englefield,nicola,ioffe}. It is immediate to show that $\phi_0=e^{\int w(x)dx}$ satisfies $H\phi_0=0$, and therefore, $\phi_0$ satisfies equation (\ref{susy}) with eigenvalue $\lambda=0$. Hence, the stationary Fokker-Planck equation admits solutions with eigenvalues $\lambda\geq0$. This version of the Fokker-Planck equation is well known \cite{polloto,anjos,ho}.

Our present purpose is to study the Fokker-Planck equation and its solutions by replacing the standard derivative  $\frac{\partial}{\partial x}$ with any of the following Dunkl-type derivatives
\begin{eqnarray}
&&D_Y\equiv\frac{\partial}{\partial x}-\frac{\mu}{x}R, \label{yan}\\
&&D_D\equiv\frac{\partial}{\partial x}+\frac{\mu}{x}-\frac{\mu}{x}R, \label{dun}\\
&&D_{CH}\equiv\frac{\partial}{\partial x}+\frac{\sigma}{x}-\frac{\mu}{x}R,\label{dun1}\\
&&D_{TP}\equiv\frac{\partial}{\partial x}+\frac{\mu}{x}-\frac{\mu}{x}R +\gamma \frac{\partial}{\partial x}R, \label{dun2}
\end{eqnarray}
which are the Yang, Dunkl, Chung-Hassanabadi and Two-Parameter Chung-Hassanabadi derivatives, respectively \cite{chung1,chung2}. Here, the parameters $\sigma$ and $\mu$ must satisfy $\sigma>1/2$ and $\mu>-1/2$ \cite{chung2}, and $R$ is the reflection operator with respect to the $x$-coordinate. Thus, the action of $R$ on any function $f(x)$ is given by $Rf(x)=f(-x)$, and therefore
\begin{equation}
 R^2=1, \hspace{3ex}\frac{\partial}{\partial x}R=-R\frac{\partial}{\partial x},\hspace{3ex}Rx=-xR, \hspace{3ex}R D_x=-D_xR\label{pro1},
\end{equation}
with $D_x$ any of the Dunkl-type derivatives (\ref{yan})-(\ref{dun2}).

We notice that the CH derivative (\ref{dun1}) is more general than those of Yang and Dunkl (\ref{yan}) and (\ref{dun}), since if we set $\sigma=0$, $D_{CH}\rightarrow D_Y$  and if we set $\sigma=\mu$, $D_{CH}\rightarrow D_D$. Thus, the more general of the first three derivatives (which are of the same kind) is the Chung-Hassanabadi derivative $D_{CH}$. Consequently, in what follows we will generalize the Fokker-Planck equation changing the spatial derivative by the CH and TP derivatives. If we substitute $D_I$, $I=CH, TP$ into equation (\ref{FP}) we obtain the time-dependent Dunkl-Fokker-Planck equation
\begin{equation}
\frac{\partial \mathcal{P}(x,t)}{\partial t} =-D_{I}\left(2w(x)\mathcal{P}(x,t)\right)+D^2_{I}\mathcal{P}(x,t) \label{DFP}.
\end{equation}
From now on, in this work we will consider the following probability density
\begin{equation}
\mathcal{P}(x,t)=e^{-\lambda t}\psi(x).
\end{equation}
Hence, if we introduce in equation (\ref{DFP}) the separable variables product for this probability density,
we obtain the Dunkl-Fokker-Planck eigenvalues equation
\begin{equation}
-D^2_{I}\psi(x)+2D_{I}(w(x)\psi(x))=\lambda\psi(x),\hspace{3ex}I=CH,\,TP.\label{DFP2}
\end{equation}
By direct calculation, we show that the second order CH and TP derivatives are given by
\begin{equation}
D_{CH}^2=\frac{\partial^2}{\partial x^2}+\frac{2\sigma}{x}\frac{\partial}{\partial x}+\frac{\sigma^2-\mu^2-\sigma}{x^2}+\frac{\mu}{x^2}R, \label{lap1}
\end{equation}
and
\begin{equation}
D_{TP}^2=(1-\gamma^2)\left(\frac{\partial^2}{\partial x^2}+\frac{2\eta}{x}\frac{\partial}{\partial x}-\frac{\eta}{x^2}+\frac{\eta}{x^2}R\right), \label{lap2}
\end{equation}
where we have defined $\eta=\frac{\mu}{1-\gamma}$. With this definition, the derivative (\ref{dun2}) takes the form
\begin{equation}
D_{TP}=\frac{\partial}{\partial x}+\frac{(1-\gamma)\eta}{x}-\frac{(1-\gamma)\eta}{x}R +\gamma \frac{\partial}{\partial x}R.\label{dun3}
\end{equation}
By substituting the operators (\ref{dun1}) and (\ref{lap1}) into equation (\ref{DFP}), we obtain the generalized Fokker-Planck equation for the CH derivative
\begin{multline}
\left[-\frac{d^2}{d x^2}-\frac{2\sigma}{x}\frac{d}{d x} - \frac{\sigma^2-\mu^2-\sigma}{x^2}-\frac{\mu}{x^2}R+2\left(\frac{d w(x) }{d x}\right)\right.\\
\left.
+2w(x)\frac{d}{d x}+\frac{2\sigma w(x)}{x}-\frac{2\mu}{x}(R w(x))R\right]\psi(x)=\lambda\psi(x).\label{DFP2CH}
\end{multline}
Similarly, for the TP derivatives (\ref{lap2}) and (\ref{dun3}) we obtain the following generalized Fokker-Planck equation
\begin{multline}
\left[-(1-\gamma^2)\left(\frac{d^2}{d x^2}+\frac{2\eta}{x}\frac{d}{d x}-\frac{\eta}{x^2}+\frac{\eta}{x^2}R\right)+2\left(\frac{dw(x)}{dx}\right)+2w(x)\frac{d}{dx}+\frac{2(1-\gamma)\eta}{x}w(x)\right.\\
\left.
-\frac{2(1-\gamma)\eta}{x}(R w(x))R+2\gamma\left(\frac{d (Rw(x)) }{d x}\right)R+2\gamma (Rw(x))\frac{d}{dx}R\right]\psi(x)=\lambda\psi(x).\label{DFP3TP}
\end{multline}
In order to obtain these two generalized Fokker-Planck equations we have used that
\begin{equation}
R(w(x)\psi(x))=w(-x)\psi(-x)=(Rw(x))(R\psi(x)).\label{parity}
\end{equation}
Notice that from equation (\ref{DFP3TP}),  $\gamma$  must be a real number such that $\gamma\in(-1,1)$. Now, we must focus on finding drift functions leading to generalized Fokker-Planck equation being exactly solvable, where $w(x)=D^{(1)}(x)/2$.
An important point to be emphasized is that so far, in deriving equations (\ref{DFP2CH}) and (\ref{DFP3TP}) we have not assumed any parity property on the superpotential function $w(x)$.
We also point out that we have preserved the name of superpotential for $w(x)$ since by setting the Dunkl parameters to vanish in our generalized Fokker-Planck equations (\ref{DFP2CH}) or (\ref{DFP3TP}), they reduce to equation (\ref{FP}) or equivalently to expression (\ref{susy}). However, in this work we will not introduce any supersymmetry or shape invariance treatment.
In the applications of the generalized Fokker-Planck equations we will solve the differential equations analytically.

We are interested in finding the definite parity
eigenfunctions. Hence, we demand that the reflection operator $R$ must commute with the operator on the left hand side of equation (\ref{DFP2}). For this goal to be achieved, the superpotential $w(x)$ must be restricted to be an odd function. Thus, as applications, in Sections 3 and 4 we obtain the eigenfunctions and energy spectrum for the centrifugal-type potential $a(a-1)/x^2$ and the harmonic oscillator plus a centrifugal-type potential.

The normalization of the wave functions of quantum mechanics with Dunkl-type derivative generalizations are given by
\begin{equation}
\int_{-\infty}^{\infty}\psi^*_\lambda(x)\psi_{\lambda'}(x)|x|^{2\sigma}dx=\delta_ {\lambda\lambda'}
\end{equation}
and
\begin{equation}
\int_{-\infty}^{\infty}\psi^*_\lambda(x)\psi_{\lambda'}(x)|x|^{2\eta}dx=\delta_ {\lambda\lambda'}.
\end{equation}
for the CH and the TP derivatives, respectively \cite{chung2,NOS4}.

\section{Generalized Fokker-Planck equation solution of the superpotential $a/x$ for the CH derivative}

By setting $w(x)=a/x$, for $a\in \mathbb{R}\backslash\{1\}$, we obtain the potential
\begin{equation}
w(x)^2+w'(x)=\frac{a(a-1)}{x^2}.
\end{equation}
We investigate the eigenfunctions of equation (\ref{DFP2CH}) which are of definite parity.\\

I) Even parity solutions. \\

These are obtained by setting $R\psi(x) =\psi(x)$ into equation (\ref{DFP2CH}), which then reduces to the differential equation
\begin{equation}
-x^2\frac {d ^2}{d x^2}f(x)+2x \left( a-\sigma \right) \frac {d}{d x}f(x) +\left( -\lambda x^2+ \left( \mu+
\sigma-1 \right)  \left( \mu+2a-\sigma \right)  \right) f(x)=0.
\end{equation}
Since $\lambda>0$, we can introduce the new variable  $u\equiv\sqrt{\lambda}x$ . Thus, this equation results to be
\begin{equation}
\frac {d ^2}{d u^2}f(u)+\frac{2\sigma-2a}{u} \frac {d}{d u}f(u) +\left( 1+ \frac{(1- \mu-
\sigma)( \mu+2a-\sigma)}{u^2}  \right) f(u)=0, \label{rp}
\end{equation}
which is a Lommel's type equation.

It is known that the Lommel's equation
\begin{equation}
v''(z)+\frac{1-2\alpha}{z}v'(z)+\left(\left( \beta\gamma z^{\gamma-1}\right)^2+\frac{\alpha^2-\nu^2\gamma^2}{z^2}\right)v(z)=0,\label{lommel}
\end{equation}
has as solutions the regular functions at the origin
\begin{equation}
v(z)=z^\alpha J_\nu(\beta z^\gamma),
\end{equation}
where $ J_\nu(z)$ are the first kind Bessel functions of order $\nu$  \cite{lebedev,nikiforov}.
By direct comparison of equations (\ref{rp}) and (\ref{lommel}), we find
\begin{equation}
\alpha=a-\sigma+\frac{1}{2}\hspace{6ex}\beta=1,\hspace{6ex}\gamma=1,\hspace{6ex}\nu=\left|a+\mu-\frac{1}{2}\right|.
\end{equation}
Thus, the regular solutions at the origin of equation (\ref{rp}) are given by
\begin{equation}
\psi(u)=C_e u^{a-\sigma+\frac{1}{2}}J_{\left|a+\mu-\frac{1}{2}\right|}(u)=C'_e x^{a-\sigma+\frac{1}{2}}J_{\left|a+\mu-\frac{1}{2}\right|}(\sqrt{\lambda}x). \label{tipo1}
\end{equation}

II) Odd parity solutions.\\

Similarly, by setting  $R\psi(x) =-\psi(x)$ into equation (\ref{DFP2CH}), we find
\begin{equation}
-x^2\frac {d ^2}{d x^2}f(x)+2x \left( a-\sigma \right) \frac {d}{d x}f(x) +\left( -\lambda x^2+ \left(
\sigma-\mu-1 \right)  \left( 2a-\mu-\sigma \right)  \right) f(x)=0,
\end{equation}
which under the change of variable $u\equiv\sqrt{\lambda}x$ transforms also into a Lommel's type equation
\begin{equation}
\frac {d ^2}{d u^2}f(u)+\frac{2\sigma-2a}{u} \frac {d}{d u}f(u) +\left( 1+ \frac{(\mu-
\sigma+1)( 2a-\mu-\sigma)}{u^2}  \right) f(u)=0. \label{rp2}
\end{equation}
Also, as in the previous case, a direct comparison with equation (\ref{lommel}) lead us to find
\begin{equation}
 \psi(u)=C_o u^{a-\sigma+\frac{1}{2}}J_{\left|a-\mu-\frac{1}{2}\right|}(u)=C'_ox^{a-\sigma+\frac{1}{2}}J_{\left|a-\mu-\frac{1}{2}\right|}(\sqrt{\lambda}x)\label{tipo2}
\end{equation}
as its regular solutions at the origin. 

Here it is important to note that the eigenfunctions obtained in equations (\ref{tipo1}) and (\ref{tipo2}) for the even and odd cases are valid for any $\lambda\in \mathbb{R}$, $\lambda>0$. Therefore, for this problem its spectrum is in the continuum.

Notice that the functions in equations (\ref{tipo1}) and (\ref{tipo2}) have arisen in the context of the generalization of the Schr\"odinger equation with Dunkl derivative \cite{chung1,hassa,NOS5}, and the analysis is finished without further investigation.
In references \cite{SCH} and \cite{chung2}, the authors imposed additional restrictions to the eigenfunctions resulting from solving the Sch\"odinger equation with Dunkl derivative, to satisfy the required parity. Our expressions (\ref{tipo1}) and (\ref{tipo2}) are solutions to the Dunkl-Fokker-Planck equation and we have imposed that $R\psi(x) =\pm \psi(x)$. However, it is known that the only Bessel functions of definite parity are those of integer order. For this reason, we must impose additional restrictions on the eigenfunctions  (\ref{tipo1}) and (\ref{tipo2}) to have the required parity.

By basic properties of Bessel functions of integer order, we have
\begin{equation}
R(J_m(x))=J_m(-x)=(-1)^mJ_m(x), \hspace{6ex}R(x^nJ_m(x))=(-1)^{n+m}J_m(x),
\end{equation}
with $n$ and $m$ integer numbers. Thus, the parity of the functions  $x^nJ_m(x)$ is given by $n+m$. Also it is known that the functions $x^nJ_m(x)$ are regular at the origin for $n=-m, -m+1, -m+2,...$

Therefore, for a given value of the parameter $a$ in the superpotential, we must impose that the subscripts of the Bessel functions  (\ref{tipo1}) and (\ref{tipo2}) take integer values. Thus, we have the following cases:\\

a) For the even parity eigenfunctions, $|a+\mu-\frac{1}{2}|=m=0,1,2,3,...$ implies that $\mu=$$\frac{1}{2}-a$,$\frac{3}{2}-a$,$\frac{5}{2}-a$,$\frac{7}{2}-a$,... or  $\mu=$ $\frac{1}{2}-a$,-$\frac{1}{2}-a$, -$\frac{3}{2}-a$,-$\frac{5}{2}-a$,..., whenever $\mu>-\frac{1}{2}$.\\

b) For the odd parity eigenfunctions, the restriction $|a-\mu-\frac{1}{2}|=m=0,1,2,3,...$ implies that $\mu$ can takes the values  $\mu=$$a-\frac{1}{2}$, $a-\frac{3}{2}$, $a-\frac{5}{2}$, $a-\frac{7}{2}$,... or  $\mu=$ $a-\frac{1}{2}$, $a+\frac{1}{2}$, $a+\frac{3}{2}$, $a+\frac{5}{2}$,...., whenever $\mu>-\frac{1}{2}$.\\

In order to make our eigenfunctions (\ref{tipo1}) and (\ref{tipo2}) regular at the origin, the powers of $x$ must be restricted to take the values $a-\sigma+\frac{1}{2}$=$-m$, $-m+1$, $-m+2$,..., and as a consequence, $\sigma=$$m+a+\frac{1}{2}$, $m+a-\frac{1}{2}$, $m+a-\frac{3}{2}$,..., whenever $\sigma>-\frac{1}{2}$.

As a particular case of the above argument, setting $a=2$, we find that $\mu=$$\frac{1}{2}$, $\frac{3}{2}$, $\frac{5}{2}$, $\frac{7}{2}$... and $\sigma=$$\frac{1}{2}$, $\frac{3}{2}$, $\frac{5}{2}$, for $m=0$,  $\sigma=$$\frac{1}{2}$, $\frac{3}{2}$, $\frac{5}{2}$, $\frac{7}{2}$, for $m=1$,... For the plots (a) and (b) in Figure 1, we have set $a=2$, $\lambda=4$, and the values of $\mu$ and $\sigma$, according to Table 1.
\begin{table}[ht]
\begin{center}
\caption{Particular eigenfunctions of the generalized Fokker-Planck equation for $w(x)=a/x$ with $\lambda=4$ and $a=2$.}
\renewcommand{\arraystretch}{1.2}
\begin{tabular}{| r | l | c |c|}\hline\hline
$\mu$& $\sigma$ & $even$&$odd$ \\ \hline \hline
$\frac{1}{2}$&$\frac{5}{2}$ &$x^0J_2(2x)$&$x^0J_1(2x)$\\ \hline
$\frac{11}{2}$ & $\frac{7}{2}$ & $x^{-1}J_7(2x)$&$x^{-1}J_4(2x)$\\ \hline
$\frac{9}{2}$ & $\frac{9}{2}$& $x^{-2}J_6(2x)$&$x^{-2}J_3(2x)$\\ \hline\hline
\end{tabular}
\renewcommand{\arraystretch}{1}

\label{table1}
\end{center}
\end{table}

\vspace{-0.2in}

\begin{figure}[ht]
 \centering
  \subfloat[]{
   \label{pares}
    \includegraphics[width=0.45\textwidth]{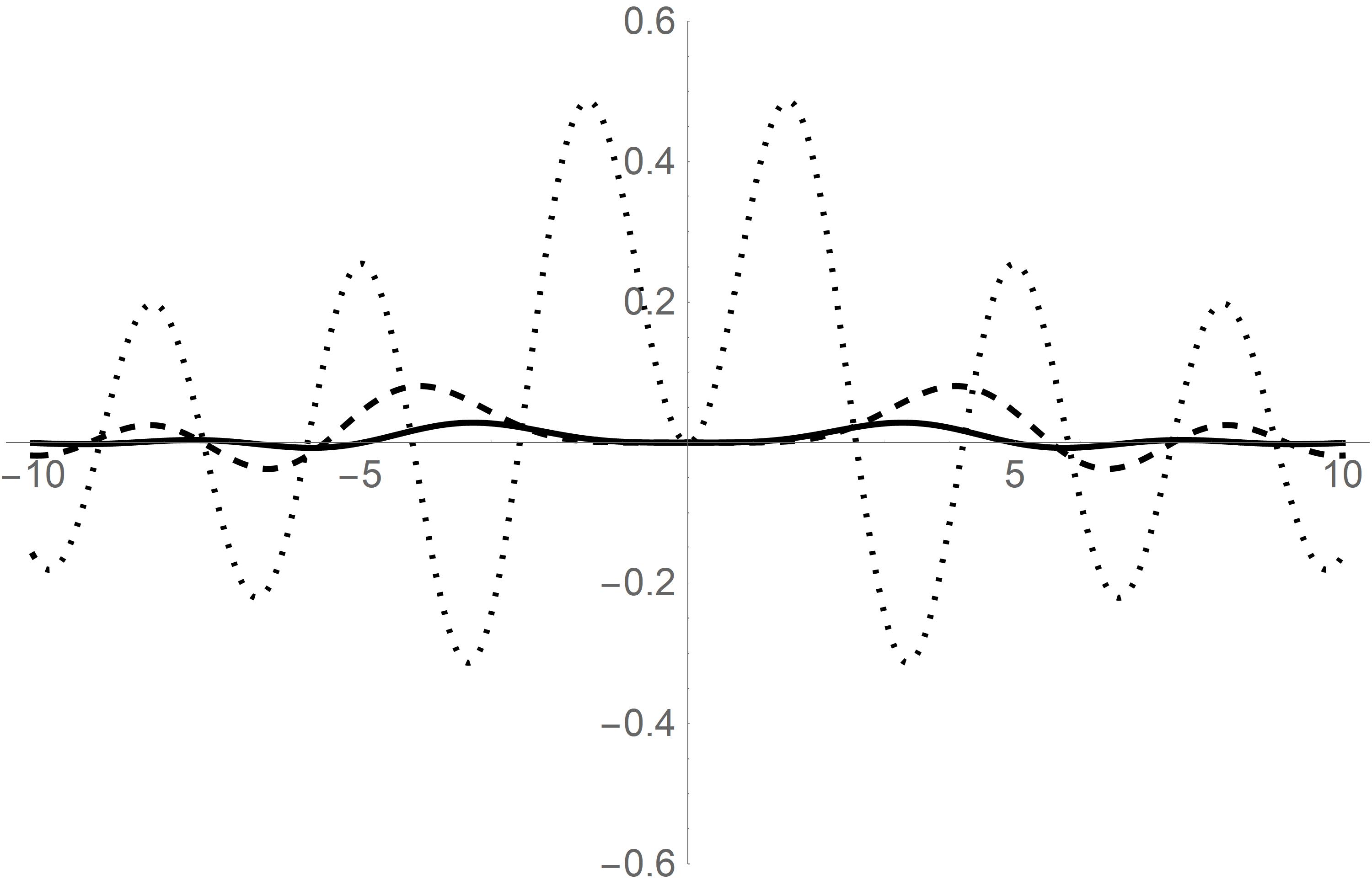}}\hspace{0.5cm}
  \subfloat[]{
   \label{impares}
    \includegraphics[width=0.45\textwidth]{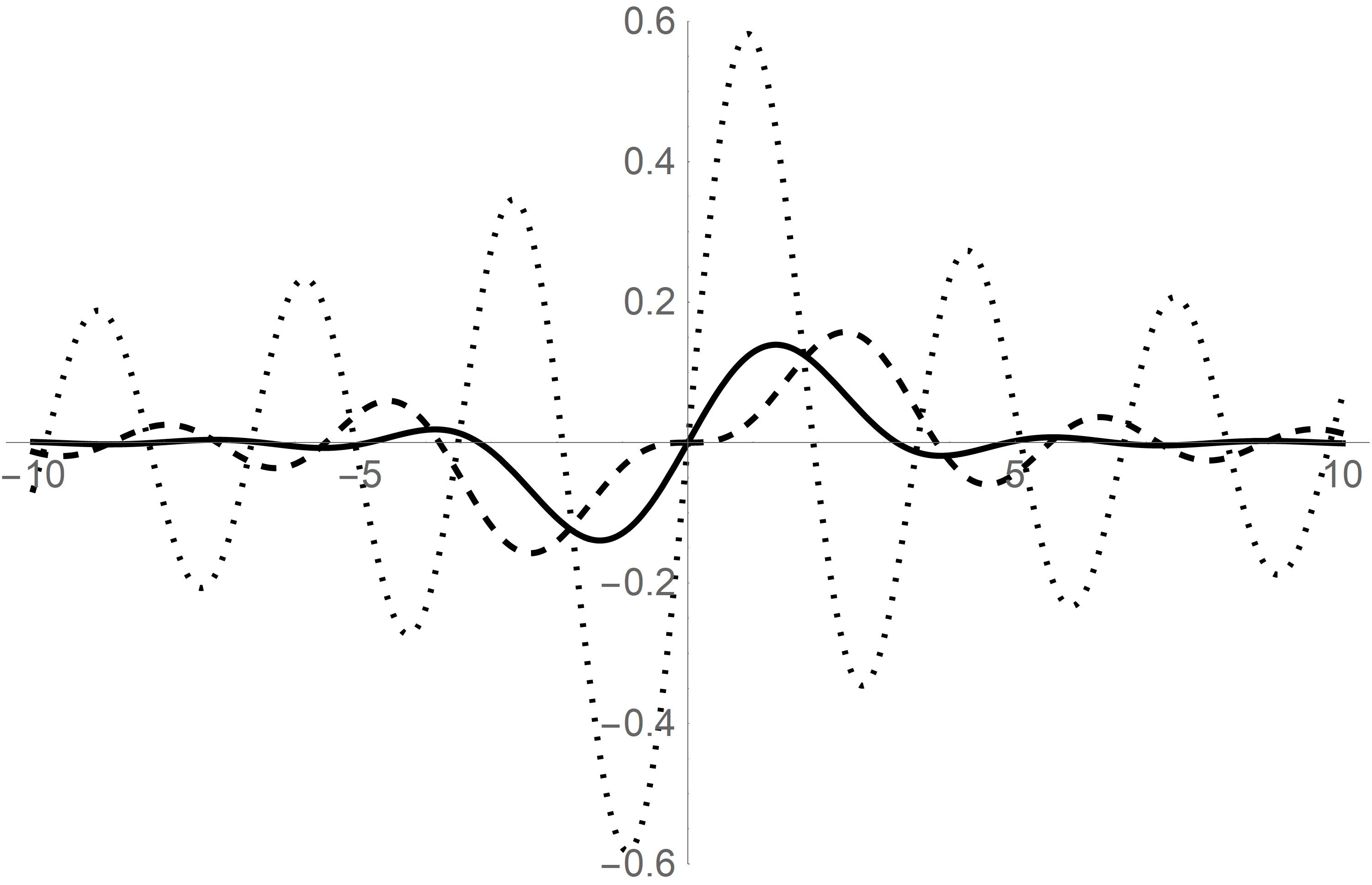}}
  \caption{\footnotesize Plots of Bessel eigenfunctions with $a=2$, $\lambda=4$. Figure (a) shows the even parity eigenfunctions $x^{0}J_2(2x)$ (dot), $x^{-1}J_7(2x)$ (dash), $x^{-2}J_6(2x)$ (solid). Figure (b) shows the odd parity eigenfunctions $x^{0}J_1(2x)$ (dot), $x^{-1}J_4(2x)$ (dash), $x^{-2}J_3(2x)$ (solid).}
\end{figure}

\section{Generalized Fokker-Planck equation solution of the harmonic oscillator plus a centrifugal-type potential for the TP derivative}

The harmonic oscillator plus a centrifugal-type potential is reproduced by the odd superpotential $w(x)=\frac{a}{x}-x$, with $a\in \mathbb{R}\backslash\{1\}$,
since in this case
\begin{equation}
w(x)^2+w'(x)=\frac {a( a-1) }{x^2}+x^2-2a-1.\label{potencial}
\end{equation}
Note in advance that we have introduced the term $a/x$ in $w(x)$ so that the probability density tends to zero as $x\rightarrow \pm \infty$ .
Thus, we will solve the generalized Fokker-Planck equation (\ref{DFP3TP}) for the TP derivative with $w(x)=\frac{a}{x}-x$ and $Rw(x)=-\frac{a}{x}+x$.\\

I) Even parity solutions.\\

These are obtained by setting $R\psi_e(x) =\psi_e(x)$ into equation (\ref{DFP3TP}) to get
\begin{multline}
{x}^{2} \left( {\gamma}^{2}-1 \right) \frac { d ^2}{d x^2}\psi_e(x) -2x \left( \gamma-1 \right)
\left(a -\eta(\gamma+1)-x^2\right) \frac { d}{ d x}\psi_e(x)\\
+\left[  \left(  \left( 4\eta+2 \right)( \gamma-1)-
\lambda \right) {x}^{2}+ \left(1- \gamma \right)  \left( 4\eta-
2 \right) a \right]\psi_e(x) =0.\label{sho1}
\end{multline}
By substituting
\begin{equation}
\psi_e(x)=e^{-\frac{x^2}{1+\gamma}}x^{\frac{2a}{1+\gamma}}g_e(x), \label{potencial2}
\end{equation}
equation (\ref{sho1}) transforms into
\begin{equation}
x \frac{d^2}{d x^2 }g_e(x)+\left(2\eta+\frac{2a}{1+\gamma}-2\frac{x^2}{1+\gamma}\right)\frac{d}{d x}g_e(x)-\frac{\lambda}{\gamma^2-1}xg_e(x)=0
\end{equation}
and introducing  a new variable $u=\frac{x^2}{1+\gamma}$ , this equation takes the form
\begin{equation}
u\frac{ d ^2}{d u^2} g_e( u)+\left(\frac{1}{2}+\eta+\frac{a}{1+\gamma}-u\right) \frac{d}{ d u}g_e(u) +\frac{\lambda}{4(1-\gamma)}g_e( u) =0.\label{sho2}
\end{equation}
It is well known that the solutions to the differential equation
\begin{equation}
z \frac {{ d}^{2}}{{d}{z}^{2}}f\left( z \right)+ \left( \alpha+1 -z\right) {\frac
{ d}{{ d}z}}f\left( z \right) + n f\left( z \right) =0, \label{laguerre}
\end{equation}
are given by the Laguerre polynomials \cite{lebedev}
\begin{equation}
f(z) =  L_n^\alpha (z), \hspace{5ex}n=0,1,2,3...,\hspace{5ex}\alpha>-1.
\end{equation}
Thus, a direct comparison between (\ref{sho2}) and (\ref{laguerre}) leads us to identify $g_e(u)=L_{n_e}^{\alpha_e}(u)$, with parameters
\begin{equation}
\alpha_e=\eta-\frac{1}{2}+\frac{a}{1+\gamma},\hspace{5ex}\hspace{5ex}n_e=\frac{\lambda}{4(1-\gamma)}.\label{lp}
\end{equation}
Therefore, for the even eigenfunctions, we obtain that the spectrum of the harmonic oscillator plus a centrifugal-type potential for the generalized Fokker-Planck equation is
\begin{equation}
\lambda=4n_e(1-\gamma).
\end{equation}
Since $\gamma\in(-1,1)$ and $n_e\geq0$ we also get that $\lambda\geq0$. The above results imply that
\begin{equation}
\psi_e(x)=C_ee^{-\frac{x^2}{1+\gamma}}x^{\frac{2a}{1+\gamma}}L_{n_e}^{\alpha_e}(x^2/(1+\gamma)), \label{solucion1}
\end{equation}
are the even parity solutions of the generalized Dunkl-Fokker-Planck equation for the harmonic oscillator plus a centrifugal-type potential.\\

II) Odd parity solutions. \\

In this case, we set $R\psi_o(x)=-\psi_o(x)$ and the generalized Fokker-Planck equation (\ref{DFP3TP}) is given by
\begin{multline}
{x}^{2} \left( {\gamma}^{2}-1 \right) \frac { d ^2}{d x^2}\psi_o(x) +2x \left( \gamma+1 \right)
\left(a +\eta(\gamma-1)-x^2\right) \frac { d}{ d x}\psi_o(x)\\
-\left[  \left( 2 ( \gamma+1)+
\lambda \right) {x}^{2}+2 \left(\gamma+1 \right)\left(\eta(\gamma-1)+a \right)\right]\psi_o(x) =0.\label{invertida2}
\end{multline}
Now, if we define
\begin{equation}
\psi_o(x)=e^{-\frac{x^2}{1-\gamma}}x^{\frac{2a}{1-\gamma}-2\eta}g_o(x),\label{potencia2}
\end{equation}
then  $g_o(x)$  must satisfy the equation
\begin{equation}
x \frac{d^2}{d x^2 }g_o(x)-\left(2\eta+\frac{2a}{\gamma-1}-2\frac{x^2}{\gamma-1}\right)\frac{d}{d x}g_o(x)-\frac{\lambda}{\gamma^2-1}xg_o(x)=0.
\end{equation}
In terms of the new variable $u=\frac{x^2}{1-\gamma}$, this differential equation takes the form
\begin{equation}
u\frac{ d ^2}{d u^2} g_o( u)+\left(\frac{1}{2}-\eta+\frac{a}{1-\gamma}-u\right) \frac{d}{ d u}g_o(u) +\frac{\lambda}{4(\gamma+1)}g_o(u) =0,\label{sho3}
\end{equation}
which can be identified with the Laguerre equation (\ref{laguerre}). Thus, in this case we obtain the following values for the parameters $\alpha_o$ and $n_o$:
\begin{equation}
\alpha_o=\frac{a}{1-\gamma}-\eta-\frac{1}{2},\hspace{5ex}\hspace{5ex}n_o=\frac{\lambda}{4(1+\gamma)},\label{lp2}
\end{equation}
and as a consequence, $g_o(u)=L_n^\alpha(u)$.
Therefore, for the odd eigenfunctions, the last equation allows us to obtain the following energy spectrum
\begin{equation}
\lambda=4n_o(1+\gamma).
\end{equation}
Similarly, in this case we have $\lambda\geq0$, provided that $\gamma\in(-1,1)$ and $n_o\geq0$. Also, the results above lead us to find
\begin{equation}
\psi_o(x)=C_oe^{-\frac{x^2}{1-\gamma}}x^{\frac{2a}{1-\gamma}-2\eta}L_{n_o}^{\alpha_o}(x^2/(1-\gamma)), \label{solucion2}
\end{equation}
as the odd eigenfunctions of the generalized Dunkl-Fokker-Planck equation for the harmonic oscillator plus a centrifugal-type potential.

It is important to notice that when $\gamma=0$, all the results obtained for this problem reduce to those reported in \cite{NOS6}, where the Fokker-Planck equation is studied with the standard Dunkl derivative.

At this stage we emphasize that to obtain the eigenfunctions (\ref{solucion1}) and (\ref{solucion2}) of the generalized Dunkl-Fokker-Planck equation  we explicitly assumed that $R\psi(x)=\pm\psi(x)$. This is why we call them even and odd parity solutions, respectively.
In what follows, as in the previous example, we must impose additional restrictions on the eigenfunctions so that they have the required parity.\\

a)  The eigenfunctions (\ref{solucion1}) are even if the exponent of $x$ is an even integer, for a given  $a\in \mathbb{R}\backslash\{1\}$. That is to say,
$\frac{2a}{1+\gamma}=2m$, for $m=1,2,3, \dots$. From this result, we find that $\gamma=\frac{a}{m}-1$. Since $-1<\gamma<1$, then  $-1<\frac{a}{m}-1<1$ or equivalently  $0<\frac{a}{m}<2$. Thus, for example, if $a=4.3$, the possible allowed values for $m$ are $m=3, 4, 5, 6,...$

In Table 2 we have displayed the first four even eigenfunctions of the generalized Dunkl-Fokker-Planck equation with $m=2$. The associated Laguerre polynomials where taken from \cite{arfken}.\\

\begin{table}[ht]
\begin{center}
\caption{First four even eigenfunctions of the generalized Dunkl-Fokker-Planck equation for the harmonic oscillator plus a centrifugal-type potential.}
\renewcommand{\arraystretch}{1.7}
\begin{tabular}{| r | l |}\hline\hline
$n_{e}$& \hspace{1cm}$\psi_e(x)$ \hspace{3cm}$\beta\equiv 1+\gamma$\hspace{1.5cm}$\alpha_e=\eta-\frac{1}{2}+\frac{a}{1+\gamma}$ \\ \hline \hline
$0$& $e^{-\frac{x^2}{\beta}}x^6$\\ \hline
$1$& $e^{-\frac{x^2}{\beta}}x^6\left(\alpha_e+1-\frac{x^2}{\beta}\right)$\\ \hline
$2$ &$ e^{-\frac{x^2}{\beta}}x^6\left(\frac{(\alpha_e+1)(\alpha_e+2)}{2}+\frac{(\alpha_e+2)x^{2}}{\beta}+\frac{(\alpha_e+2)x^4}{(2\alpha_e+4)\beta^2}\right) $\\ \hline
$3$& $e^{-\frac{x^2}{\beta}}x^6\left(\frac{(\alpha_e+1)(\alpha_e+2)(\alpha_e+3)}{6}-\frac{(\alpha_e+2)(\alpha_e+3)x^2}{2\beta}+\frac{(\alpha_e+2)(\alpha_e+3)x^4}{(2\alpha_e+4)\beta^2}-\frac{(\alpha_e+2)(\alpha_e+3)x^6}{(2\alpha_e+4)(3\alpha_e+9)\beta^3}\right)$\\ \hline\hline
\end{tabular}
\renewcommand{\arraystretch}{1}
\label{2 table}
\end{center}
\end{table}
b) First,  let us also remember that $\eta\equiv \frac{\mu}{1-\gamma}$. Since $-1<\gamma<1$, then it follows that $0<1-\gamma<2$. On the other hand, the eigenfunctions (\ref{solucion2}) have odd parity if the exponent of $x$ is an odd integer. That is to say $\frac{2a}{1-\gamma}-2\eta =\frac{2(a-\mu)}{1-\gamma}$$=2m+1$, for $m=0,1,2,3, \dots$. Hence, $\frac{2(a-\mu)}{2m+1}=1-\gamma$. This means that $0<\frac{2(a-\mu)}{2m+1}<2$ and $a>\mu$. As an example, if $a=4.3$ and $\mu=0.6$, then, the allowed values for $m$ are $m=2, 3, 4, 5, \dots$.
Thus, if we choose $m=2$, the power of $x$ is 5. The table for the first odd eigenfunctions turns out to  be the same as Table 2, except that we must change the power 6 by 5, $\beta$ by $\beta=1-\gamma$ and $\alpha_e$ by $\alpha_o=\frac{a}{1-\gamma}-\eta-\frac{1}{2}$.

\begin{figure}[ht]
 \centering
  \subfloat[]{
   \label{paresLag}
    \includegraphics[width=0.45\textwidth]{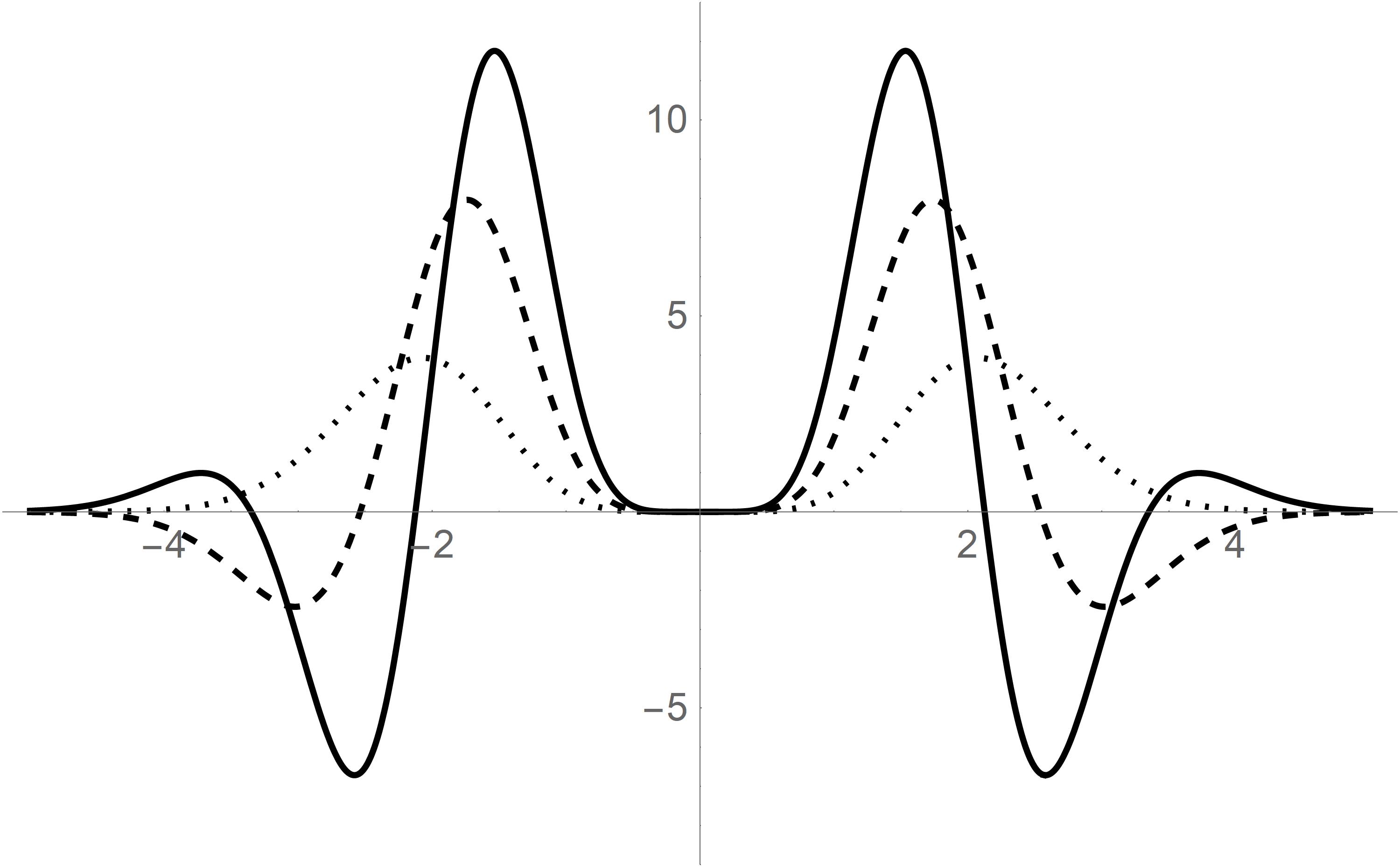}}\hspace{0.5cm}
  \subfloat[]{
   \label{imparesLag}
    \includegraphics[width=0.45\textwidth]{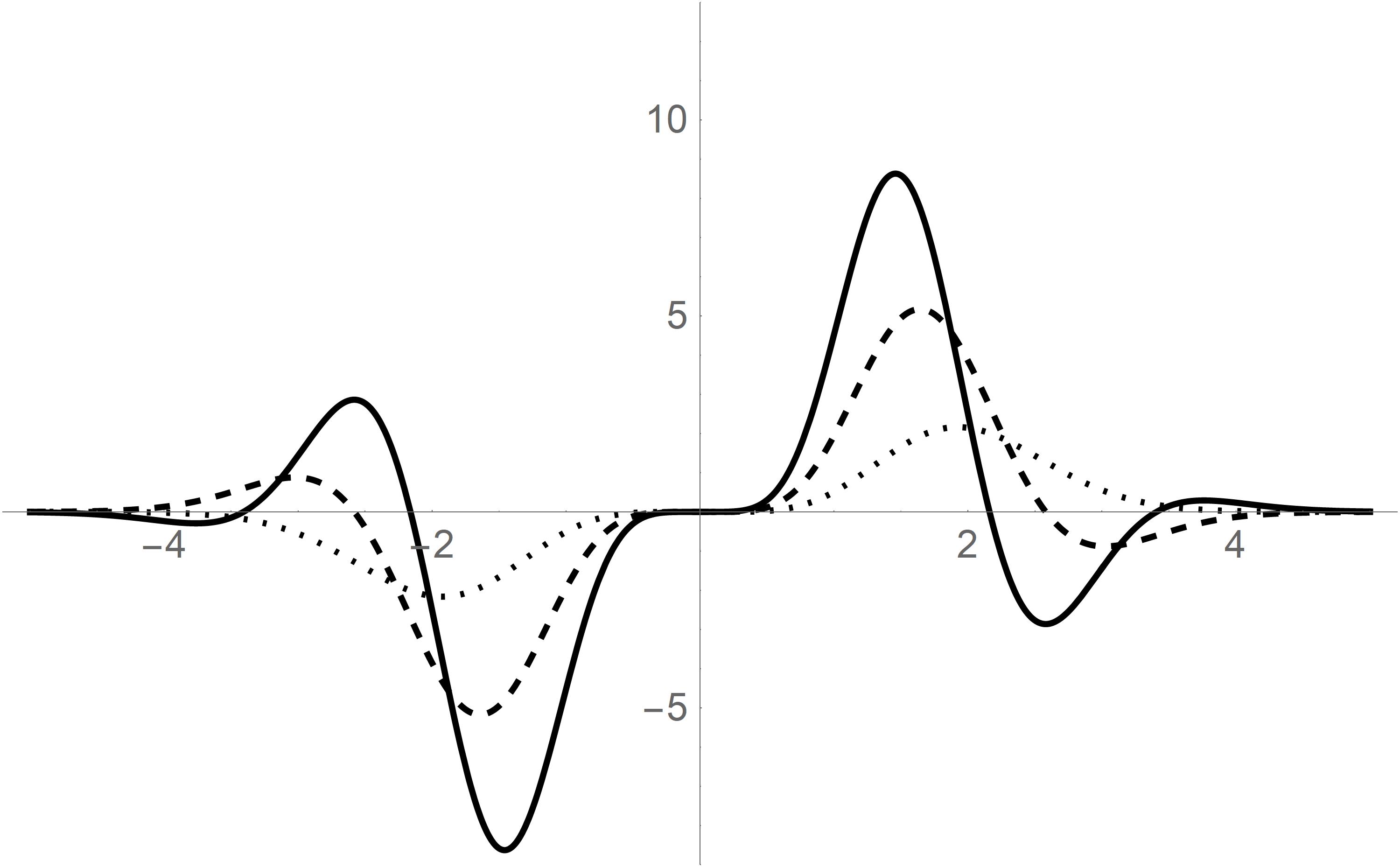}}
  \caption{\footnotesize Eigenfunctions of the harmonic oscillator plus a centrifugal-type potential with $a=4.3$, $\mu=0.6$. Figure (a) shows the even parity eigenfunctions for $\alpha_e=3.5$, $\gamma=0.43$ for $n_e=0$ (dot), $n_e=1$ (dash), $n_e=2$ (solid). Figure (b) shows the odd parity eigenfunctions for $\alpha_o=2$, $\gamma=-0.48$ for $n_o=0$ (dot), $n_o=1$ (dash), $n_o=2$ (solid).}
\end{figure}

In Figure (2) we have plotted the first three even and odd eigenfunctions, respectively. The even parity eigenfunctions are plotted with $a=4.3$, $\mu=0.6$, $\alpha_e=3.5$, $\gamma=0.43$ and $m=3$, while the odd parity eigenfunctions are plotted with $a=4.3$, $\mu=0.6$, $\alpha_o=2$, $\gamma=-0.48$ and $m=2$. As can be seen, all these eigenfunctions have the expected parity.

\section{Concluding Remarks}
The Fokker-Planck equation has been generalized in two different ways. The first generalization was in terms of the Chung-Hassanabadi derivative, which generalizes the Dunkl and Yang derivatives. The second one was in terms of the Two-Parameter Chung-Hassanabadi derivative. The generalized Fokker-Planck equations introduced in the present work for these Dunkl-type derivatives are as general as possible. However, to have definite parity solutions, the superpotential $w(x)$ must be chosen as an odd function.

In order to provide some applications of our general results, we obtained the eigenfunctions and the energy spectrum of the superpotential $a/x$ and the harmonic oscillator plus a centrifugal-type potential in a closed form. The potential $a(a-1)/x^2$  was studied by solving the Fokker-Planck equation generalized by the Chung-Hassanabadi derivative in terms of the Bessel functions. Similarly, the exact solutions of the harmonic oscillator plus a centrifugal-type potential were computed by solving the Fokker-Planck equation generalized by the Two-Parameter Chung-Hassanabadi derivative in terms of the Laguerre polynomials.

\section*{Acknowledgments}
We would like to thank the anonymous referees for their valuable comments to improve our work.\\

This work was partially supported by SNII-M\'exico, COFAA-IPN, EDI-IPN, and CGPI-IPN Project Numbers $20230633$ and $20230732$, and CONAHCYT grant CB-2017-2018-A1-S-30345.

\end{document}